# Non-Hermitian Complex Coupling for Magnetic Resonance Imaging

Siyong Zheng (郑思涌),[1,4] Maopeng Wu (吴茂鹏),[2] Zhonghai Chi (池中海),[1] Zewen Chen (陈泽炆),[3] Chwee Teck Lim,[4,5] Qian Zhao (赵乾),[2,†] and Ji Zhou (周济)[1,†]

[1] State Key Laboratory of New Ceramics and Fine Processing, School of Materials Science and Engineering, Tsinghua University, Beijing 100084, China
[2] State Key Laboratory of Tribology in Advanced Equipment, Department of Mechanical Engineering, Tsinghua University, Beijing 100084, China
[3] Weixian College, Tsinghua University, Beijing, 100084, China
[4] Department of Biomedical Engineering, National University of Singapore, 9 Engineering Drive 1, Singapore, 117575, Singapore
[5] Institute for Health Innovation and Technology (iHealthtech), National University of Singapore, Singapore, 117599, Singapore

**ABSTRACT**. Strong coupling in wave-based systems often causes level repulsion, leading to mode splitting and reduced response at the target frequency. This problem is pronounced in magnetic resonance imaging (MRI), where strong mutual inductance between a receive coil (RC) and a metamaterial (MM) degrades $B_1$ performance. Here, we introduce a non-Hermitian complex-coupling decoupling strategy based on a dual-resonator model. By engineering a phase delay in the coupling pathway, an imaginary coupling component is created, driving the system from the PT-symmetric to the anti-PT-symmetric phase and enabling eigenmode degeneracy without added dissipation. Implemented through a high-permittivity ceramic layer, this mechanism restores single-mode resonance in the MM–RC system and suppresses frequency splitting. Simulations show a ~14-fold $B_1$ enhancement compared with the strongly repulsive regime. This passive, compact, and hardware-compatible approach offers a general route for coupling control in electromagnetic, acoustic, optical, and quantum systems.

## I. INTRODUCTION.

Strong coupling is one of the key mechanisms governing energy exchange and modal interactions in a broad range of wave systems [1]. When two resonant units operate at nearby frequencies and exhibit substantial mutual inductance or near-field interaction, the system enters a strong-coupling regime, where the eigenfrequency spectrum displays a characteristic mode splitting. This phenomenon reflects energy exchange and interference between the two resonators and represents a universal physical behavior observed from quantum systems to electromagnetic devices [2–4]. However, strong coupling also introduces an unavoidable side effect—level repulsion [5]. The frequency splitting of the hybridized modes weakens the system's response at the target operating frequency, thereby reducing signal sensitivity or energy-transfer efficiency [6].

In engineering applications, level repulsion induced by strong coupling becomes particularly problematic. In electromagnetic devices, acoustic cavities, optical resonators, and magnetic resonance imaging (MRI) systems, near-field interactions between components often shift the resonance away from the desired frequency, making it difficult to simultaneously maintain high coupling efficiency and high sensitivity [7–9]. This behavior can be summarized as a "strong-coupling, strong-repulsion" effect. Established decoupling strategies rely on geometric tuning and conventional circuit-based approaches, such as spacing adjustment or overlap optimization, to suppress excessive mutual coupling. For example, in nonlinear MMs designed for parallel MRI, the coupling strength was regulated by adjusting the separation between a MM and the receive coil to avoid resonance detuning [10]. These methods are intuitive and effective. However, they tend to compromise structural compactness and increase circuit complexity, which is particularly problematic in MR scanners where the available space is extremely limited. Moreover, adding active circuitry to an otherwise passive device also sacrifices the advantages of passive operation.

Recently, non-Hermitian physics has offered a new perspective on this long-standing challenge [11]. By introducing unbalanced gain and loss, or more generally, complex-valued coupling terms, non-Hermitian system can exhibit spectral behaviors unattainable in conventional Hermitian systems—most notably, level attraction [12–14]. In contrast to level repulsion, level attraction drives the two eigenfrequencies toward degeneracy. Under specific conditions, the system reaches an exceptional point (EP) at which the two modes share the same resonance frequency and imaginary part (i.e., zero gain/loss contrast), enabling eigenmode degeneracy and dramatically enhanced responses. Importantly, this mechanism can be realized through phase engineering of complex coupling without adding extra dissipative elements, thereby providing a pathway toward passive decoupling in strongly coupled systems [15].

Building upon this concept, this work proposes a non-Hermitian complex-coupling-driven decoupling strategy for dual resonators (hereafter referred to as the "complex-coupling strategy"). By introducing a controlled phase delay in the coupling pathway, an imaginary coupling component is generated, enabling level attraction and spectral degeneracy under strong-coupling conditions. This work first establishes a general non-Hermitian dual-resonator model and analyze how the coupling strength, loss difference, and coupling phase collectively shape eigenvalue evolution, revealing the continuous transition from parity-time (PT) symmetry to anti-PT-symmetry. This work then maps this physical mechanism onto the strong mutual coupling between a MM and a receive coil in MR scanners and propose an engineering implementation based on phase modulation using high-permittivity dielectric materials. This enables passive coil decoupling and enhanced sensitivity without introducing additional loss.

This work show that the proposed complex-coupling strategy effectively suppresses frequency splitting in MR scanners and enhances the $B_1$ sensitivity near the Larmor frequency by approximately 14-fold.

†Contact author: zhaoqian@tsinghua.edu.cn

†Contact author: zhouji@tsinghua.edu.cn

Compared to spatial-separation approaches, the proposed method requires no active components or lossy elements, remains structurally compact, and offers excellent compatibility, manufacturability, and scalability. More importantly, the strategy is not limited to MRI; it represents a general non-Hermitian decoupling mechanism applicable to optical, acoustic, and quantum coupled systems, providing a new theoretical foundation and implementation paradigm for controlling and designing strongly coupled multi-physics systems.

## II. NON-HERMITIAN PHYSICAL MODEL OF THE DUAL-RESONATOR SYSTEM

### A. Level repulsion in real-coupling systems

To elucidate the physical origins of level repulsion and level attraction in strongly coupled systems, this work begins with a simplified model consisting of two mutually coupled resonators. The two resonant units possess intrinsic resonance frequencies $\omega_1$ and $\omega_2$, as well as dissipation terms $\Gamma_1$ and $\Gamma_2$. They are connected through a coupling coefficient $k$, forming a dual-resonator system in which energy can be exchanged. The dynamics of the system can be expressed as

$$i\frac{da_1}{d_t} = \omega_1 a_1 - i\Gamma_1 a_1 + k a_2 \quad (1)$$
$$i\frac{da_2}{d_t} = \omega_2 a_2 - i\Gamma_2 a_2 + k a_1$$

where $a_i$ ($i = 1, 2$) denotes the resonating amplitudes. The two resonators in this model are placed sufficiently close to each other, with a separation much smaller than the electromagnetic wavelength at their resonance frequencies. The dynamical equation exhibits typical PT symmetry, with its non-Hermitian nature arising from the dissipation of the resonators. The effective Hamiltonian corresponding to Eq. (1) can be written as

$$\hat{H} = \begin{bmatrix} \omega_1 - i\Gamma_1 & k \\ k & \omega_2 - i\Gamma_2 \end{bmatrix}$$
$$= \begin{bmatrix} \omega_1 & k \\ k & \omega_2 \end{bmatrix} - i\begin{bmatrix} \Gamma_1 & 0 \\ 0 & \Gamma_2 \end{bmatrix} \quad (2)$$

When $\omega_1 = \omega_2 = 0$, the analytical solution of the eigenfrequencies is given by:

$$E_\pm = -i\bar{\Gamma} \pm \sqrt{k^2 - \Delta\Gamma^2/4} \quad (3)$$

†Contact author: zhaoqian@tsinghua.edu.cn

†Contact author: zhouji@tsinghua.edu.cn

Here, $\bar{\Gamma} = (\Gamma_1 + \Gamma_2)/2$ and $\Delta\Gamma = \Gamma_2 - \Gamma_1$. When $\Delta\Gamma = 0$, the system is Hermitian, and the square root term in Eq. (3) becomes real. In this case, the real parts of the two eigenfrequencies are split by $2k$, producing the characteristic mode splitting shown in Fig. 1b, i.e., level repulsion. Physically, as the coupling strength increases, the resonance modes of the two resonators move further apart, and the energy distributes into two orthogonal hybridized modes. This phenomenon is widely observed in strongly coupled optical cavities, microwave resonators, and near-field MM systems, where the real parts of the eigenfrequencies separate monotonically with increasing $k$.

When the loss contrast $\Delta\Gamma \neq 0$ is taken into account, the system Hamiltonian becomes non-Hermitian, and a competition arises between the coupling term and the dissipation terms. As shown in Fig. 1c, the frequency splitting in the real parts of the eigenvalues gradually decreases as $\Delta\Gamma$ increases, and eventually vanishes at $\Delta\Gamma = 2k$. At this point, the system reaches the EP, where the two modes share the same resonance frequency and imaginary part, resulting in eigenmode degeneracy. This behavior indicates that adjusting the dissipation balance can drive the system from level repulsion to level attraction, offering a potential route toward passive decoupling in strongly coupled systems.

### B. Complex-coupling model with phase delay

In conventional coupled systems, the coupling coefficient $k$ is typically treated as a real number, corresponding to energy exchange between the two resonators without any phase delay. When an additional phase delay $\varphi$ is introduced into the coupling pathway, the coupling term becomes complex in the form of $ke^{i\varphi}$ (Fig. 1d). In this case, the effective Hamiltonian of the system can be written as

$$\hat{H} = \begin{bmatrix} \omega_1 - i\Gamma_1 & \kappa \\ \kappa & \omega_2 - i\Gamma_2 \end{bmatrix} =$$
$$\begin{bmatrix} \omega_1 & \text{Re}(\kappa) \\ \text{Re}(\kappa) & \omega_2 \end{bmatrix} - i\begin{bmatrix} \Gamma_1 & -\text{Im}(\kappa) \\ -\text{Im}(\kappa) & \Gamma_2 \end{bmatrix} \quad (4)$$

Here, the complex coupling $\kappa$ consists of a real-valued coupling strength $k$ and an imaginary component introduced by the phase factor $e^{-i}$. As shown in Fig. 1e, when the coupling phase $\varphi$ increases from 0 to $\pi/2$, the system undergoes a continuous transition from PT symmetry to anti-PT symmetry. The splitting in the real parts of the eigenfrequencies decreases progressively and eventually disappears at $\varphi = \pi/2$, exhibiting the characteristic level-attraction behavior. Along the $\Delta\Gamma$ direction, the system shows the typical evolution between the gain–loss–balanced PT-symmetric phase and the PT-broken phase.

When the resonance-frequency mismatch $\Delta\omega$ is further considered (Fig. 1f), there exists a parameter regime in which the real parts of the two modes coincide while the imaginary parts reach their maximum separation, indicating that the system resides in the anti-PT-symmetric phase. The eigenmode degeneracy in this regime enables a single-peak response to be maintained even under strong-coupling conditions, providing the physical foundation for passive decoupling in later sections.

## III. IMPLEMENTATION IN MRI

### A. Strong coupling and mode splitting in MRI

In MR scanners, MMs are often placed around the receive coil (RC) to enhance the local magnetic field and thereby improve signal sensitivity (Fig. 2a) [16–21]. However, when the MM has a comparable size to the coil and the separation between them is small, the system enters a strong-coupling regime. In this case, the resonance frequencies of the MM and the RC are typically designed near the Larmor frequency $\omega_0$, and the two elements interact through near-field mutual inductance, forming a canonical dual-resonator system (Fig. 2b). Under these conditions, the effective Hamiltonian of the system takes the same form as Eq. (1), with the coupling strength $k$ determined by the mutual inductance between the coil and the MM. Full-wave simulations show that as the size of the MM increases or the gap decreases, $k$ grows rapidly and significant mode splitting emerges in the spectrum (Fig. 2c). The resulting frequency splitting reduces the magnetic-field response at $\omega_0$, leading to degraded sensitivity—a manifestation of the "strong-coupling, strong-repulsion" effect.

Conventional decoupling methods typically rely on adjusting the physical separation between components [22–24]. Although intuitive and easy to implement, such approaches face significant limitations in MR scanners: the available space is highly constrained due to the limited bore size and compact geometry, particularly in head or extremity imaging. As a result, it is difficult to achieve effective decoupling without compromising the structural integrity of the system.

†Contact author: zhaoqian@tsinghua.edu.cn

†Contact author: zhouji@tsinghua.edu.cn

On the other hand, the PT-symmetry (Eq. (3)) indicates that the spectral characteristics of the system are governed by the loss contrast $\Delta\Gamma$. As $\Delta\Gamma$ increases, the system undergoes a transition from the PT-symmetry to the PT-broken phase, reaching an EP at which the two modes become degenerate. In MRI-related simulations, this transition can be induced by adding extra resistive loading to the MM rings to increase dissipation, thereby restoring mode degeneracy (Fig. 2d). However, this approach serves primarily as a theoretical demonstration: the introduction of resistors leads to energy dissipation and a substantial reduction in the magnetic-field enhancement factor, degrading signal sensitivity and increasing system noise. Thus, although loss tuning via resistive loading can validate PT-symmetric models, it is impractical for real MR scanners. Achieving decoupling under strong-coupling conditions, without introducing additional dissipation and while preserving the compact geometry of the system, remains a key challenge.

### B. Principle of passive decoupling driven by complex coupling

Based on the non-Hermitian framework discussed above, we propose a complex-coupling – driven decoupling strategy for dual resonators. The core idea is to introduce an imaginary component of the coupling through phase-delay modulation, enabling level attraction and eigenmode degeneracy without adding any additional dissipation.

A high-permittivity ceramic layer is inserted between the MM and the RC, introducing a phase delay $\phi$ along the coupling pathway (Fig. 3a). This delay can be regarded as a complex phase shift accumulated during energy exchange between the two resonators, effectively replacing the coupling term in Eq. (4) with $ke^{i\varphi}$. The resulting complex coupling introduces asymmetric phase interference, driving the system into the anti-PT-symmetry and enabling eigenmode degeneracy without altering the intrinsic dissipation.
This mechanism can be understood as a modification of the phase condition governing energy exchange between the two resonators. When $\varphi \approx \pi/2$, the imaginary component of the coupling dominates the system's behavior, switching the interaction between the two modes from repulsion to attraction. The system can be abstracted as two resonators coupled through a phase delay, and the eigenvalue spectrum exhibits a continuous evolution from mode splitting to mode degeneracy as the coupling phase increases (Fig. 3b).

Simulation results show that the magnetic-field frequency response evolves from two distinct modes to a single mode as the dielectric constant increases, corresponding to a continuous reduction of the effective wavelength in the dielectric. At permittivity is approximately 320, the two modes fully merge into one (Fig. 3c). Beyond this point, the two eigenmodes reappear.

Further tuning the ring-end capacitance, which corresponds to adjusting the resonance-frequency mismatch $\Delta\omega$, reveals a similar mode degeneracy (Fig. 3d). The frequency separation between the two modes gradually decreases and eventually vanishes when the capacitance lies in the range of 17.5~22.5 pF, within which only a single-mode magnetic-field frequency response is observed. As the capacitance increases beyond 22.5 pF, the difference between the two modes becomes pronounced again, and the spectrum once more exhibits two distinct resonances. This behavior is fully consistent with the theoretical eigenvalue evolution shown in Fig. 1e,f.

## IV. PARAMETER OPTIMIZATION AND DESIGN CRITERIA FOR MODE DEGENERACY

To achieve optimal decoupling performance, it is necessary to determine the appropriate combination of ceramic thickness and dielectric constant. The overall design workflow is illustrated in Fig. 4a: starting from the MRI operating frequency and geometric constraints, the available space is first identified, followed by a parameter sweep of the ceramic properties to locate the degeneracy point. The modal frequency difference exhibits a pronounced minimum as the ceramic thickness varies (Fig. 4b). When the thickness is approximately 10.4 mm, the frequency difference reaches its minimum, corresponding to complete mode degeneracy.

Figures 4c and 4d show the evolution of the real and imaginary parts of the eigenfrequencies with thickness. At the degeneracy point, the real parts coincide while the imaginary parts reach their maximum separation, indicating that the system enters the anti-PT-symmetric phase. This parameter-selection criterion provides a practical guideline for designing MRI MMs: when the dielectric thickness and permittivity satisfy $\varphi \approx \pi/2$ and $|\Delta\omega| \leq 2k$, mode degeneracy is achieved.

Under the optimized parameter conditions, the MM implementing the complex-coupling–driven dual-resonator decoupling strategy exhibits a markedly higher magnetic-field strength in the MRI system compared with MMs operating in the "strong-coupling, strong-repulsion". As shown in Fig. 5a, the conventional configuration produces a split $B_1$ frequency response with two peaks due to level repulsion, resulting in a magnetic-field strength at the operating frequency of only 0.1× (vs. BC). In contrast, the MM operating under the anti-PT-symmetric dual-resonator decoupling strategy restores a single-peak resonance, yielding a ~14-fold enhancement in $B_1$ strength (vs. RC–MM).

These results demonstrate that, under strong-coupling conditions, the complex-coupling–driven strategy effectively suppresses frequency splitting and significantly improves sensitivity at the Larmor frequency. The approach requires no modification to existing MRI hardware, involves no active components or resistive dissipation, and is fully compatible with commercial MR scanners. Physically, the mechanism relies on phase-induced complex coupling that drives the system into the anti-PT-symmetric phase, allowing modal resonance to be maintained and enabling passive decoupling together with substantial signal enhancement.

## V. CONCLUSIONS

In summary, the coherence of dissipative coupling has been shown to govern the transition between symmetric and antisymmetric phase in non-Hermitian system. Building on a non-Hermitian model of two coupled resonators, this work systematically elucidates the physical evolution from level repulsion induced by real-valued coupling to level attraction enabled by complex-valued coupling. By introducing a controlled phase delay to construct complex coupling, mode degeneracy and $B_1$-field enhancement are achieved.

The complex coupling driven dual resonator decoupling strategy developed here enables passive decoupling and sensitivity enhancement for MRI MMs, offering flexible control over coupling phase, resonator frequency, and dissipation. By engineering the dielectric constant, ceramic thickness, and resonator frequencies to realize the antisymmetric phase, mode degeneracy can be maintained at the operating frequency even under strong-coupling conditions. Compared with the strong-repulsion regime of conventional MRI MMs, the proposed approach yields a 14-fold enhancement in $B_1$ strength.

†Contact author: zhaoqian@tsinghua.edu.cn

†Contact author: zhouji@tsinghua.edu.cn

This study provides a compelling example of applying non-Hermitian physics to electromagnetic-wave engineering. The physical mechanism and design strategy demonstrated here may be further extended to acoustic, optical, and quantum-information systems for coupling control and mode management, opening new avenues for designing strongly coupled multi-physics platforms.

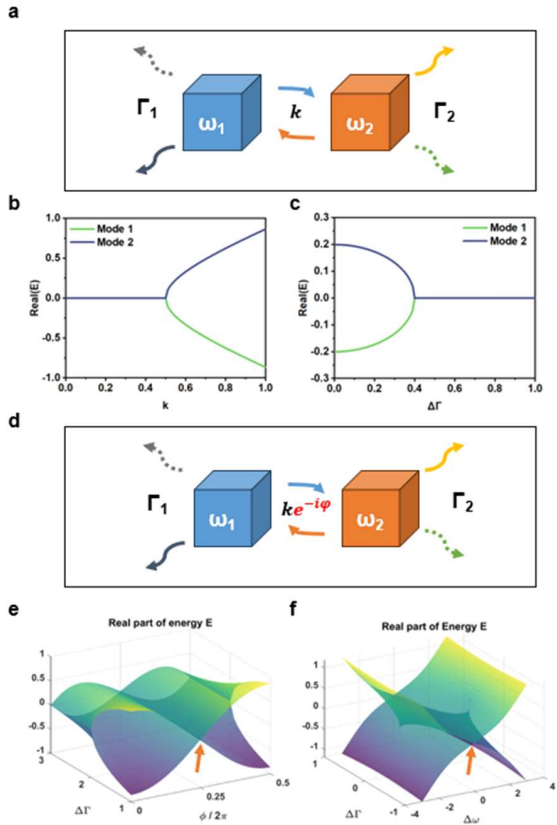

FIG 1. Non-Hermitian physical model of the dual-resonator system. (a) Schematic of two coupled resonators with coupling strength $k$ and individual dissipation rates $\Gamma_{1,2}$. (b) Real parts of the eigenfrequencies as a function of the coupling strength $k$, showing mode splitting and level repulsion. (c) Real parts of the eigenfrequencies as a function of the loss contrast $\Delta\Gamma$, illustrating mode degeneracy and the transition from the PT-symmetric to the PT-broken phase. (d) Schematic of introducing a phase delay $\varphi$ between the resonators, forming a complex coupling $\kappa$. (e) Evolution of the eigenvalue spectrum as the phase $\varphi$ increases, demonstrating the continuous transition from real-valued coupling to complex-valued coupling and finally to mode degeneracy. (f) Eigenvalue characteristics as a function of the resonance-frequency mismatch $\Delta\omega$, corresponding to the modal-degeneracy region in the anti-PT-symmetric phase.

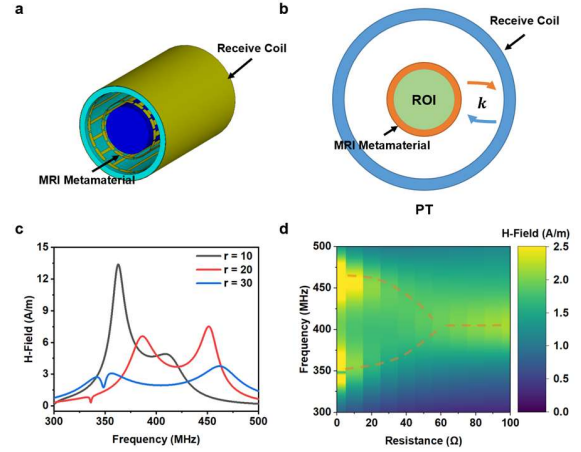

FIG 2. Strong coupling between MRI metamaterials and receive coils, and loss-induced decoupling. (a) Schematic of a conventional MRI metamaterial integrated with a receive coil, where strong near-field coupling is established. (b) Equivalent model abstracted from (a), consisting of two coupled resonators with coupling strength $k$. (c) Magnetic-field–frequency response, showing increasing mode splitting as the radius (r) of the MRI metamaterial increases, indicative of level repulsion under strong coupling. (d) By tuning the ring-end resistance to adjust the loss contrast $\Delta\Gamma$, the real parts of the eigenfrequencies become degenerate, enabling decoupling in the PT-broken phase.

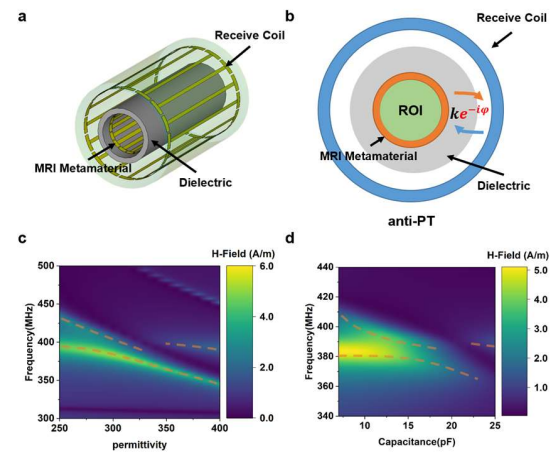

FIG 3. Implementation of the complex-coupling–driven dual-resonator decoupling strategy in MRI. (a)

†Contact author: zhaoqian@tsinghua.edu.cn

†Contact author: zhouji@tsinghua.edu.cn

Anti-PT-symmetric decoupling scheme for MRI, in which a high-permittivity ceramic layer introduces a phase delay $\varphi$ over a short distance. (b) Equivalent model abstracted from (a), consisting of two coupled resonators with a complex coupling strength $\kappa$. (c) Magnetic-field frequency responses as a function of permittivity. When permittivity is approximately 320, a $\pi/2$ phase shift is achieved, leading to mode degeneracy. (d) Magnetic-field frequency responses as a function of the tuning capacitance (corresponding to the resonance-frequency mismatch $\Delta\omega$), consistent with the eigenvalue-degeneracy condition predicted by the theoretical model.

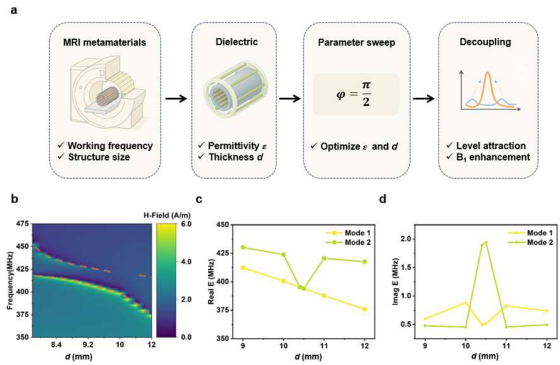

FIG 4. Parameter optimization and design criteria for mode degeneracy. (a) Flowchart of the optimization procedure for design parameters and imaging performance, illustrating the pathway from system constraints to identification of the degeneracy point. (b) Relationship between ceramic thickness (d, unit: mm) and modal frequency difference, showing a minimum at a thickness of approximately 10.4 mm, corresponding to mode degeneracy. (c) Real parts of the eigenfrequencies as a function of ceramic thickness, where the two modes coincide at the degeneracy point. (d) Imaginary parts of the eigenfrequencies as a function of ceramic thickness, indicating that the imaginary-part separation reaches a maximum in the anti-PT-symmetric phase.

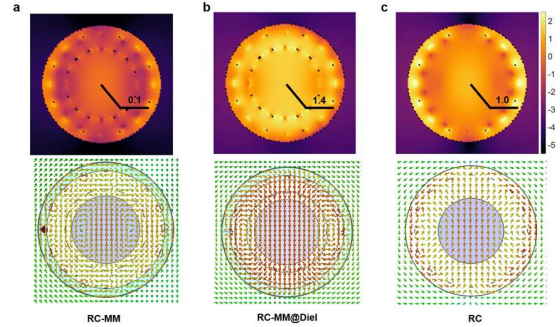

FIG 5. Performance validation of the complex-coupling–based decoupling strategy in the MR scanner (unit: lg V/m). (a) Receive coil with MRI metamaterial. (b) Dual-resonator decoupling strategy driven by complex coupling. (c) Receive coil only.

## ACKNOWLEDGMENTS
This work is supported by the National Key Research & Development Program of China (No. 2023YFB3811400), National Natural Science Foundation of China (No. 52273296) and National Natural Science Foundation of China (No. 52332006).

## METHODS
The numerical solutions and operations of "Non-Hermitian Physical Model of the Dual-Resonator System" were completed and plotted in MATLAB (MathWorks®). Numerical simulations are performed using the frequency-domain solver of CST Microwave Studio 2020 package. In open boundary situation, the simulation is excited with the circularly polarized plane wave. To acquire the magnetic field distribution, an H-Field monitor is set at the interested frequency. When the MRI metamaterial (MM) is placed in body coil (birdcage coil, BC), the simulation is excited with two discrete ports orthogonally placed in space and 90° difference in the phase. The BC was tuned to 400.0 MHz. All simulations are normalized to 1.0 W accepted power.

The geometric configuration, resonant behavior, and electromagnetic interactions of the MRI MM follow the theoretical formulation and modeling framework introduced in previous work [16].

†Contact author: zhaoqian@tsinghua.edu.cn

†Contact author: zhouji@tsinghua.edu.cn

†Contact author: zhaoqian@tsinghua.edu.cn

†Contact author: zhouji@tsinghua.edu.cn